\documentclass[fleqn,10pt]{wlscirep}
\usepackage[subrefformat=simple]{subcaption}

\usepackage[utf8]{inputenc}
\usepackage[T1]{fontenc}
\usepackage{multirow, booktabs, tabularx, longtable}
\usepackage{makecell}
\usepackage{url}
\usepackage{tikz}
\usepackage{comment}

\usepackage{amsmath} 
\usepackage[super]{nth}
\usepackage{multirow, booktabs, tabularx, longtable}
\usepackage{multirow, makecell}
\usepackage{xurl}
\usepackage{caption}
\usepackage{xfp}
\usepackage{color}
\usepackage{tabularray}
\UseTblrLibrary{booktabs}

\usepackage{filecontents}

\DefTblrTemplate{caption}{supplementary}{
  \DefTblrTemplate{caption-tag}{supplementary}{Supplementary Table \thetable}%
  \DefTblrTemplate{caption-sep}{supplementary}{.}%
  \centering
  \UseTblrTemplate{caption-tag}{supplementary}%
  \UseTblrTemplate{caption-sep}{supplementary}
  \UseTblrTemplate{caption-text}{default}\par
}
\DefTblrTemplate{firsthead}{supplementary}{\UseTblrTemplate{caption}{supplementary}}
\DefTblrTemplate{middlehead}{supplementary}{\UseTblrTemplate{caption}{supplementary}}
\DefTblrTemplate{lasthead}{supplementary}{\UseTblrTemplate{caption}{supplementary}}
\NewTblrTheme{supplementary}{
  \SetTblrStyle{caption-tag}{font=\bfseries}
  \SetTblrTemplate{firsthead}{supplementary}
  \SetTblrTemplate{middlehead}{supplementary}
  \SetTblrTemplate{lasthead}{supplementary}
}
\DeclareCaptionFormat{bfdot}{\textbf{#1}.~#3}
\definecolor{light-gray}{HTML}{FFFFFF}
\definecolor{light-blue}{HTML}{F0F0F0} 

\usepackage{tabularray}
\usepackage[hang,flushmargin,symbol]{footmisc}

\DeclareMathAlphabet\mathbfcal{OMS}{cmsy}{b}{n}
\DeclareSymbolFont{operators}   {OT1}{cmr} {m}{n}
\DeclareSymbolFont{letters}     {OML}{cmm} {m}{it}
\DeclareSymbolFont{symbols}     {OMS}{cmsy}{m}{n}

\usepackage{soul}
\usepackage{color}

\usepackage[section]{placeins}

\RequirePackage[explicit]{titlesec}
\usepackage{hyperref}
\titleformat{name=\section,numberless}
  {\large\sffamily\bfseries}
  {}
  {0em}
  {\leavevmode\MakeLinkTarget[section]{}\ignorespaces#1}
  []  
\ExplSyntaxOn\makeatletter
\def\ttl@straight@i#1[#2]#3{%
  \tl_if_empty:nTF {#2}
   {\NR@gettitle{#3}}
   {\NR@gettitle{#2}}
    \gdef\ttl@savemark{\csname#1mark\endcsname{#3}}%
  \let\ttl@savewrite\@empty
  \def\ttl@savetitle{#3}%
  \gdef\thetitle{\csname the#1\endcsname}%
  \if@noskipsec \leavevmode \fi
  \par
  \ttl@labelling{#1}{#2}%
  \ttl@startargs\ttl@straight@ii{#1}{#3}}
\ExplSyntaxOff\makeatother

\usepackage{zref-clever}
\usepackage{zref-xr}
\usepackage{subcaption}
\usepackage{zref-titleref}
\usepackage{graphicx}
\usepackage{xcolor}
\usepackage[colorlinks=true, allcolors=blue]{hyperref}
\usepackage{xr}

\zexternaldocument{supp_info}
\zcsetup{cap=true,nameinlink}

\zcLanguageSetup{english}{
  type = suppsection ,
    Name-sg = Supplementary Section ,
    name-sg = supplementary Section ,
    Name-pl = Supplementary Sections ,
    name-pl = supplementary Sections ,
  type = suppfigure ,
    Name-sg = Supplementary Figure ,
    name-sg = supplementary Figure ,
    Name-pl = Supplementary Figures ,
    name-pl = supplementary Figures ,
  type = supptable ,
    Name-sg = Supplementary Table ,
    name-sg = supplementary Table ,
    Name-pl = Supplementary Tables ,
    name-pl = supplementary Tables ,
}

\usepackage{bm}
\usepackage{float}
\usepackage{physics}
\usepackage{datetime}

\usepackage[normalem]{ulem}
\newcommand\blueout{\bgroup\markoverwith
{\textcolor{blue}{\rule[0.5ex]{2pt}{0.8pt}}}\ULon}

\title{Scaling laws of human mobility persist during extreme floods}

\author[1,2,3,4*]{Simone Loreti}
\author[5]{Markus Schl\"{a}pfer}
\author[1,2]{Andreas Paul Zischg}

\affil[1]{University of Bern, Institute of Geography, Bern, 3012, Switzerland}
\affil[2]{University of Bern, Oeschger Centre for Climate
 Change Research, Mobiliar Lab for Natural Risks, Bern, 3012, Switzerland}
\affil[3]{University of Geneva, Global Studies Institute,  Geneva, 1205, Switzerland}
\affil[4]{University of Geneva, Department of Computer Science, Carouge, 1227, Switzerland}
\affil[5]{Columbia University, Department of Civil Engineering and Engineering Mechanics, New York, 10027, New York, USA}

\affil[*]{To whom correspondence should be addressed. E-mail: \textcolor{blue}{simone.loreti@unige.ch}
\newline Date: \today
}

\begin{abstract}

\vspace{-0.5cm}

Although a number of studies have investigated human mobility patterns during natural hazards, mechanistic models that capture mobility dynamics under large-scale perturbations, such as extreme floods, remain scarce. Leveraging mobile phone data and building upon 
\textcolor{black}{recent insights into universal mobility patterns,}
we assess whether the general structure of 
\textcolor{black}{population flows}
persists during the extreme floods that struck Emilia-Romagna, Italy, in 2023.
Our analysis reveals that the relationship between visitor density, distance, and visitation frequency 
remains robust even under extreme flooding conditions.
To disentangle the effects of distance and visitation frequency, we define two aggregated visitor densities: the marginal density over frequency and the aggregated density over distance. We find that the marginal density over frequency exhibits a time-invariant power-law exponent, indicating resilience to flooding disturbances. In contrast, the aggregated density over distance displays more complex behavior: an exponential decay over biweekly periods and a power-law decay over a monthly interval. We propose that the observed power-law emerges from the superposition of exponential distributions across shorter timescales.
These findings provide new insights into human mobility scaling laws under extreme perturbations, highlighting the robustness of visitation patterns and suggesting avenues for improved mechanistic modeling during natural disasters.

\end{abstract}

\begin{document}

% ARXIV SUBMISSION

% ARXIV SUBMISSION

\maketitle
\thispagestyle{empty}

\section*{Introduction}
\label{Introduction} 

Flood is the most common \cite{Alderman2012,Mavrouli2022} and deadliest type of natural disaster worldwide \cite{Cvetković2024}. 
While the mortality rate from floods has declined over time 
\cite{Jonkman2024},
and the global occurrence has decreased since the early 2000s 
\cite{Cvetković2024,Najibi2018},
the economic toll remains substantial 
\cite{Cvetković2024}
(albeit exhibiting a mixed trend 
\cite{Coronese2019,Cvetković2024,Geiger2020,Newman2023,Paprotny2018}.
Furthermore, floods continue to exert significant social impacts 
\cite{Aznar-Crespo2021}, 
including threats to public health 
\cite{Agonafir2023,Alderman2012,Yari2024}
and disruptions to mobility 
\cite{Jafari2023,Rebally2021,Watson2022},
as well as physical impacts, such as damage to critical infrastructure 
\cite{Nirandjan2024,Pant2018},
like transportation networks 
\cite{Jafari2023,Rebally2021,Watson2022}.
Floods also cause considerable environmental damage 
\cite{Arrighi2024,Zhang2024}.
To address these challenges, sustained research and mitigation efforts are necessary 
\cite{Khan2023,Kreibich2022,Mishra2022}.
Numerous studies have explored the properties of human mobility during natural hazards 
\cite{Lu2012,Lu2016,Song2014,Wang2014,Wang2016,Yabe2019},
including floods 
\cite{Coleman2023,Loreti2025,Ma2024,Rajput2023,Yabe2020,Yuan2022},
and the resilience of infrastructure against floods. However, universal mechanistic models capable of faithfully reproducing (i) the entire dynamics of disruption in interconnected critical infrastructure, such as transportation ones, and (ii) the human mobility trajectories during and after floods, remain elusive 
\cite{Li2022,Rathnayaka2024,Tang2024,Yabe2022a,Yabe2022b,Yabe2020,Yang2024,Zhai2024}.
Schläpfer et al. 
\cite{Schläpfer2021} proposed a universal visitation law based on large-scale mobility data from diverse urban contexts, including Boston, Lisbon, Singapore, Dakar, and Abidjan. However, their analysis focused on steady-state conditions and did not account for disruptions caused by natural hazards such as floods, earthquakes, or storms. 
Motivated by the search for universal, mechanistic models of human mobility under perturbations, 
this study investigates whether similar scaling behaviors persist during extreme events. Specifically, we analyze mobility data from the 2023 flood in the Emilia-Romagna region of Italy \cite{Zaghi2024, Valente2023, Valente2025, Satriano2024, Pulvirenti2024, Portoraro2025, Montalti2024, DeFeudis2025, Cremonini2024, Arrighi2024, Ascione2025}, using anonymized 
\textcolor{black}{and aggregated}
mobile phone records provided by the Italian telecommunications company TIM. We focus on 
\textcolor{black}{the scaling behavior of visitor density with distance and travel frequency,}
as well as on aggregate quantities related to it, in the context of a large-scale natural disaster.

\section*{Results}
\label{Results}

\subsection*{Visitor density}

\begin{figure}[htp]
    \setlength{\lineskip}{0pt}
    \centering
    \includegraphics[width=\textwidth]{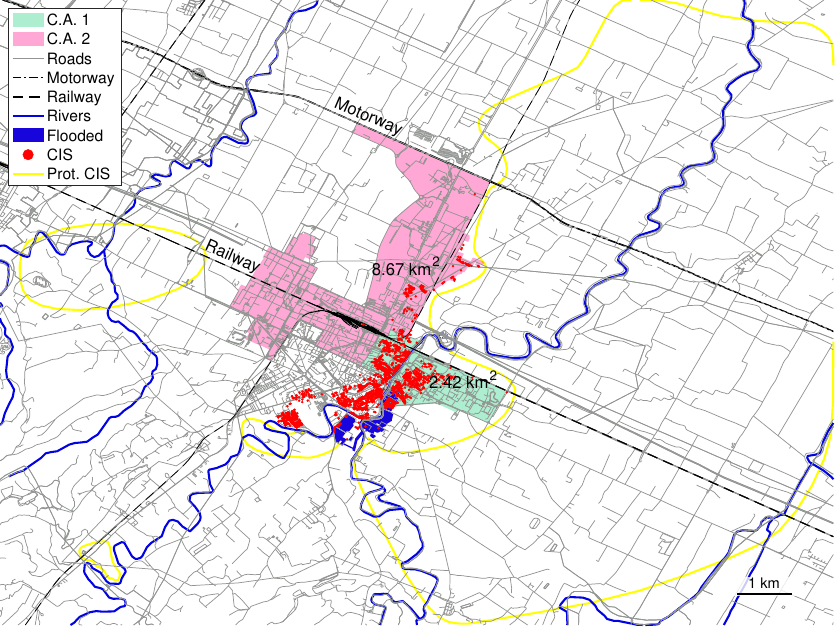} % Fig_1_CIS_PROTECTED
    \caption{
    Map of Faenza, Italy, showing the two census areas (C.A.) under examination, the flooded areas, and the locations of households that applied for the \quotes{Immediate Support Contribution} (CIS). 
    \textcolor{black}{The yellow boundaries indicate areas where CIS data density is so low that privacy protection measures are required. This figure is presented in accordance with the authorization granted by the CIS data provider, the Unione della Romagna Faentina.}
    \textcolor{black}{The map is centered at latitude $44^\circ 17' 46.54''\,\mathrm{N}$ and longitude $11^\circ 53' 41.46''\,\mathrm{E}$.}
    }
    \label{map}
\end{figure}

\begin{figure}[htp]
    \setlength{\lineskip}{0pt}
    \centering
    \includegraphics[width=1\textwidth]{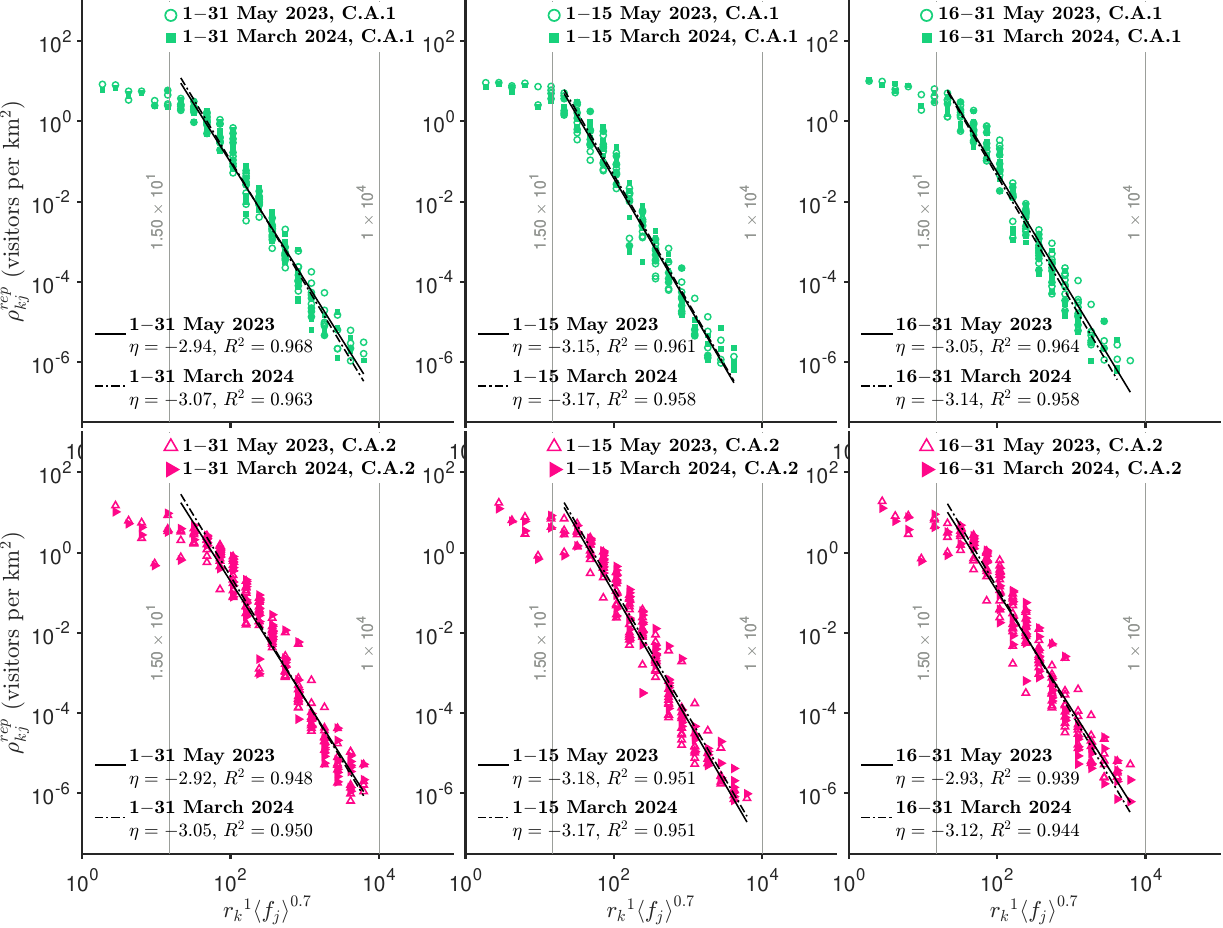}
    \caption{
    The representative visitor density, $\rho_{kj}^{\mathrm{rep}}$ (visitor density per unit area and per unit frequency) versus the product 
    \textcolor{black}{$r_k \langle f_j \rangle^{0.7}$.}
    }
    \label{visitor_density}
\end{figure}

\noindent
\textcolor{black}{Our study focuses on Faenza, one of the towns most affected by the Emilia-Romagna flood, which occurred during two major rainfall events: the first took place approximately between the \nth{1} and \nth{4} of May --- 
though specific dates vary slightly across sources, including 
1$-$3 May \cite{Cremonini2024},
1$-$4 May \cite{Arrighi2024},
2$-$3 May \cite{Pulvirenti2024}, 
and 2$-$4 May \cite{Valente2025}.
The second, more intense and geographically widespread, occurred between the \nth{16} and \nth{18} of May 2023, with sources reporting 16$-$17 May \cite{Cremonini2024} and 16$-$18 May \cite{Valente2025,Pulvirenti2024,Arrighi2024}.}
In particular, we 
\textcolor{black}{analyse}
two administrative census areas, shown in \cref{map}, which we refer to throughout the article as Census Area 1 (C.A.1) and Census Area 2 (C.A.2).
Census Area 1, located in the central part of Faenza, is crossed by a river that overflowed during the flood. This area was heavily inundated, and the numerous requests for assistance from local residents attest to the severity of the damage and provide insight into the spatial extent of the flooding. Requests for support were submitted under the emergency aid program \quotes{Contributo di immediato sostegno ai nuclei familiari} (CIS) and are represented as dots in \cref{map}. 
Census Area 2 includes part of the city center as well as an industrial zone extending to one of Italy’s main motorways, which connects northern cities (e.g., Milan) with southern regions along the Adriatic coast. This area was largely unaffected by the flood. 
For each census area, TIM provided data on the number of visitors by distance and visit 
\textcolor{black}{frequency bins}
over biweekly and monthly periods of observation, covering both the flood month (May 2023) and another entire month nearly a year later (March 2024). 
\\
\\
\textcolor{black}{Adapting the approach in \cite{Schläpfer2021} to these data (see \nameref{Methods}),}
we calculate the representative visitor density, $\rho_{kj}^{\mathrm{rep}}$ 
\textcolor{black}{(number of visitors per unit area and per unit frequency)}
for each census area and period of observation. 
\textcolor{black}{Note, however, that the data at hand have been pre-aggregated by the provider in a way that does not allow for direct comparison with the trajectory-based results in \cite{Schläpfer2021}. For instance, visitation frequencies here include several} 
\textcolor{black}{entries}
\textcolor{black}{by the same individuals}
\textcolor{black}{within}
\textcolor{black}{a single day}
\textcolor{black}{and individuals who just travel through without stopping for an activity}
\textcolor{black}{(and who}
\textcolor{black}{can not be filtered out}
\textcolor{black}{— See \nameref{Methods} \quotes{\nameref{Minimum Stay Time}}}
\textcolor{black}{).}
The representative visitor density 
\textcolor{black}{for our study}
is illustrated 
\textcolor{black}{in}
\cref{visitor_density} as a function of a variable defined as the product of the distance $r_k$ and the midpoint frequency $\langle f_j \rangle$ raised to the power of 0.7, i.e., $r_k \langle f_j \rangle^{0.7}$. 
That relationship is captured by the following power-law:
\begin{equation}
\label{rho_kj_rep}
    \rho_{kj}^{\mathrm{rep}} = a_1 \left( r_k \langle f_j \rangle^{0.7} \right) ^{\eta}
\end{equation}
The top row of \cref{visitor_density} (green points) corresponds to region C.A.1, and the bottom row (pink points) corresponds to region C.A.2. Each column represents a different time period: 1$-$31 March 2023, 1$-$15 May 2023, and 16$-$31 May 2023. For each plot, data from both 2023 and 2024 are shown with corresponding power-law fits and $R^2$ values, demonstrating the decay of visitor density with the increasing weighted distance, $r_k \langle f_j \rangle^{0.7}$. 
Considering all the census areas and observation periods, we notice that the values in \zcref{parameters_table_1} yield an average of $\langle \eta \rangle = -3.10$, with a standard deviation of $\sigma_{\eta} = 0.10$. The corresponding low coefficient of variation, i.e. $\frac{\sigma_{\eta}}{\abs{\langle \eta \rangle}} \approx 3.23\%$, suggests that $\eta$ in \cref{rho_kj_rep} is relatively stable across the different census areas and observation periods.
We can therefore write 
$
    \rho_{kj}^{\mathrm{rep}}
    \textcolor{black}{\propto}
    \left( r_k \langle f_j \rangle^{0.7} \right) ^{-3.1}
$.

\begin{table}[H]
\centering
\caption{Model parameters and total number of 
\textcolor{black}{unique users (visitors)}
for different time ranges and Census Areas (C.A.).}
\label{parameters_table_1} 

\begin{subtable}[t]{\textwidth}
\centering
\begin{tabular}{lccccc}
\toprule
\textbf{Dates} & \textbf{C.A.} & $\bm{a_1}$ & $\bm{\eta}$ & $\bm{R^2}$ & \textbf{N} \\
\midrule
1—31 May 2023    & 1 & 4.86 & $-$2.94 & 0.97 & $70,101$ \\
1—15 May 2023    & 1 & 4.88 & $-$3.15 & 0.96 & $47,199$ \\
16—31 May 2023   & 1 & 4.84 & $-$3.05 & 0.96 & $49,614$ \\
1—31 March 2024  & 1 & 5.16 & $-$3.07 & 0.96 & $72,813$ \\
1—15 March 2024  & 1 & 5.00 & $-$3.17 & 0.96 & $49,260$ \\
16—31 March 2024 & 1 & 4.92 & $-$3.14 & 0.96 & $53,452$ \\
\bottomrule
\end{tabular}
\end{subtable}

\vspace{1em}

\begin{subtable}[t]{\textwidth}
\centering
\begin{tabular}{lccccc}
\toprule
\textbf{Dates} & \textbf{C.A.} & $\bm{a_1}$ & $\bm{\eta}$ & $\bm{R^2}$ & \textbf{N} \\
\midrule
1—31 May 2023    & 2 & 5.13 & $-$2.92 & 0.95 & $168,695$ \\
1—15 May 2023    & 2 & 5.36 & $-$3.18 & 0.95 & $109,172$ \\
16—31 May 2023   & 2 & 4.91 & $-$2.93 & 0.94 & $107,599$ \\
1—31 March 2024  & 2 & 5.51 & $-$3.05 & 0.95 & $198,005$ \\
1—15 March 2024  & 2 & 5.46 & $-$3.17 & 0.95 & $120,098$ \\
16—31 March 2024 & 2 & 5.37 & $-$3.12 & 0.94 & $136,949$ \\
\bottomrule
\end{tabular}
\end{subtable}
\end{table}

\noindent
In addition to summarizing the fitted parameters, \zcref{parameters_table_1} provides the total number of 
\textcolor{black}{unique individuals}
$N$ who visited 
\textcolor{black}{each}
census areas during different periods of observation. We can then measure the percentage difference in the number of visitors between 2023 and 2024. 
For Census Area $1$ (C.A.$1$), the percentage change of the number of visitors between May $2023$ and March $2024$, i.e. $\Delta N = \left( \frac{N_{2024} - N_{2023}}{N_{2023}} \right) \times 100$, is 
\textcolor{black}{$3.87\%$}
for the entire month. In the first part of the month ($1$—$15$), there was 
\textcolor{black}{an increase} 
in the number of visitors equal to 
\textcolor{black}{$4.37\%$}
, and in the second part of the month ($16$—$31$), there was an increase of 
\textcolor{black}{$7.74\%$.}
For Census Area $2$ (C.A.$2$), the percentage change in visitors between May $2023$ and March $2024$ is 
\textcolor{black}{$\Delta N = 17.37\%$}
for the entire month. The first part of the month ($1$—$15$) saw an increase in the number of visitors equal to 
\textcolor{black}{$10.01\%$}
, while the second part of the month ($16$—$31$) experienced a more significant increase of 
\textcolor{black}{$27.28\%$.}
Despite these fluctuations in $N$, 
the scaling behavior --- describing how visitor density decays with both distance and visitation frequency --- remains remarkably stable. 
This robustness suggests that the underlying scaling pattern in \cref{rho_kj_rep} persists even during extreme flood events.
\textcolor{black}{Despite differences in the underlying data aggregation,}
our observed scaling 
\textcolor{black}{pattern}
aligns with the law reported by Schläpfer et al. (2021) \cite{Schläpfer2021}, which captures the combined influence of travel distance and visitation frequency. They reported an exponent of approximately \(\eta_{\text{Schläpfer}} \approx -2\), consistent with our combined frequency scaling of \(\langle f_j \rangle^{0.7 \langle \eta \rangle} = \langle f_j \rangle^{-2.17}\)
\textcolor{black}{, while}
we find a significantly steeper distance decay 
\(\sim r_k^{-3.1}\)
\textcolor{black}{versus}
\(\sim r_k^{-2}\)
\textcolor{black}{.}
\textcolor{black}{This difference may be explained by the pre-aggregation of our data, inherently giving more weight to individuals who live nearby and repeatedly pass through the census areas within a single day without necessarily stopping for an activity.}
\textcolor{black}{For example, during two weeks in 2023, around one-quarter of 
\textcolor{black}{travels} each week had a stay time of less than 15 minutes, suggesting a passerby behavior (see \zcref{Supp_Fig_1}). This was consistent across the last week of the flood (23$-$30 May 2023) and a week after it receded (6$-$13 June), with a substantial portion of these short-stay 
\textcolor{black}{visits also associated with short-distance trips.}
}

\subsection*{Aggregated visitor densities}

\begin{figure}[htp]
    \setlength{\lineskip}{0pt}
    \centering
    \includegraphics[width=0.9\textwidth]{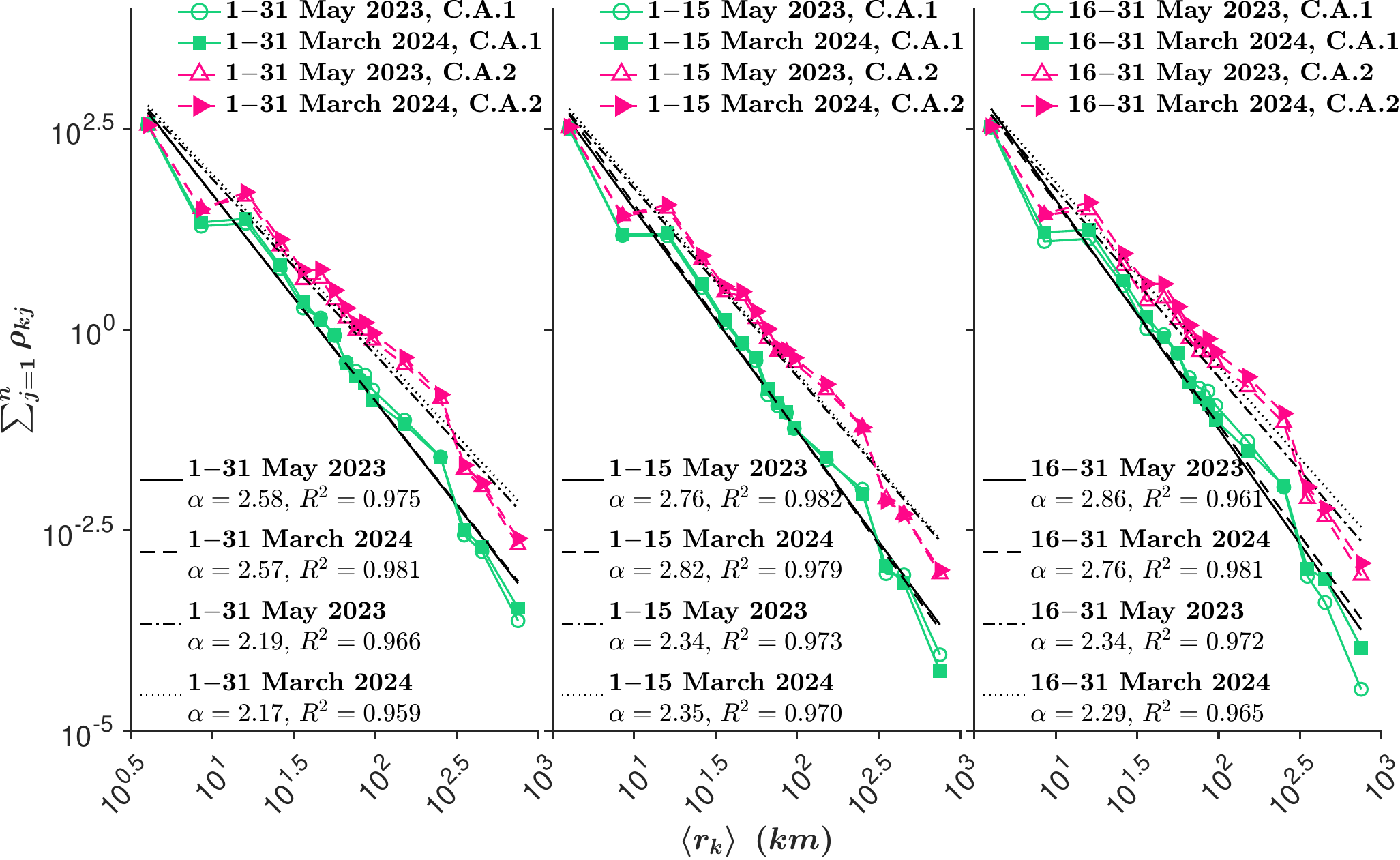}
    \caption{
    Marginal visitor density (over frequency) for distance bin $\mathcal{D}_k$, as a function of mean visit distance  $\langle r_k \rangle$.
    }
    \label{sum_rho_over_f}
\end{figure}

\noindent
\textcolor{black}{In addition to the detailed distance-frequency distribution, we assess the impact of the flooding on the visitor density marginalized over frequency and over distance.}
Please note that although this marginalization logic applies to both the raw density $\rho_{kj}$ and its normalized form $\rho_{kj}^{\mathrm{rep}}$, the interpretation of the aggregated quantities differs due to the normalization over frequency bin widths, $\frac{1}{f_{j+1} - f_j}$. Therefore, we prefer to use the non-normalized form $\rho_{kj}$, as it still represents the total visitor density.
From \cref{sum_rho_over_f}, we observe that the marginal density of visitors (over frequency) for each distance bin $\mathcal{D}_k$ follows a power law:
\begin{equation}
\label{marginal_density}
    \sum_{j=1}^n \rho_{kj} = a_2 \langle r_k \rangle ^ {\lambda}.
\end{equation}
Examining the parameter values of \cref{marginal_density} across all the observation periods (see \cref{parameters_table_2}), we find that the spatial decay of the marginal visitor density with respect to distance is, on average, steeper in Census Area~1 ($\langle \lambda \rangle \approx -2.68$) than in Census Area~2 ($\langle \lambda \rangle \approx -2.23$). However, 
\textcolor{black}{the $\lambda$ values for comparable periods across the two years and within the same Census Area remain essentially unchanged for the full-month periods ($1$—$31$), and quite stable across the biweekly intervals ($1$—$15$ and $16$—$31$). This suggests that the flooding event --- which mainly affected C.A.~1 --- had no appreciable effect on the distance decay of the mobility flows.}

\begin{figure}[htp]
    \setlength{\lineskip}{0pt}
    \centering
    \includegraphics[width=0.9\textwidth]{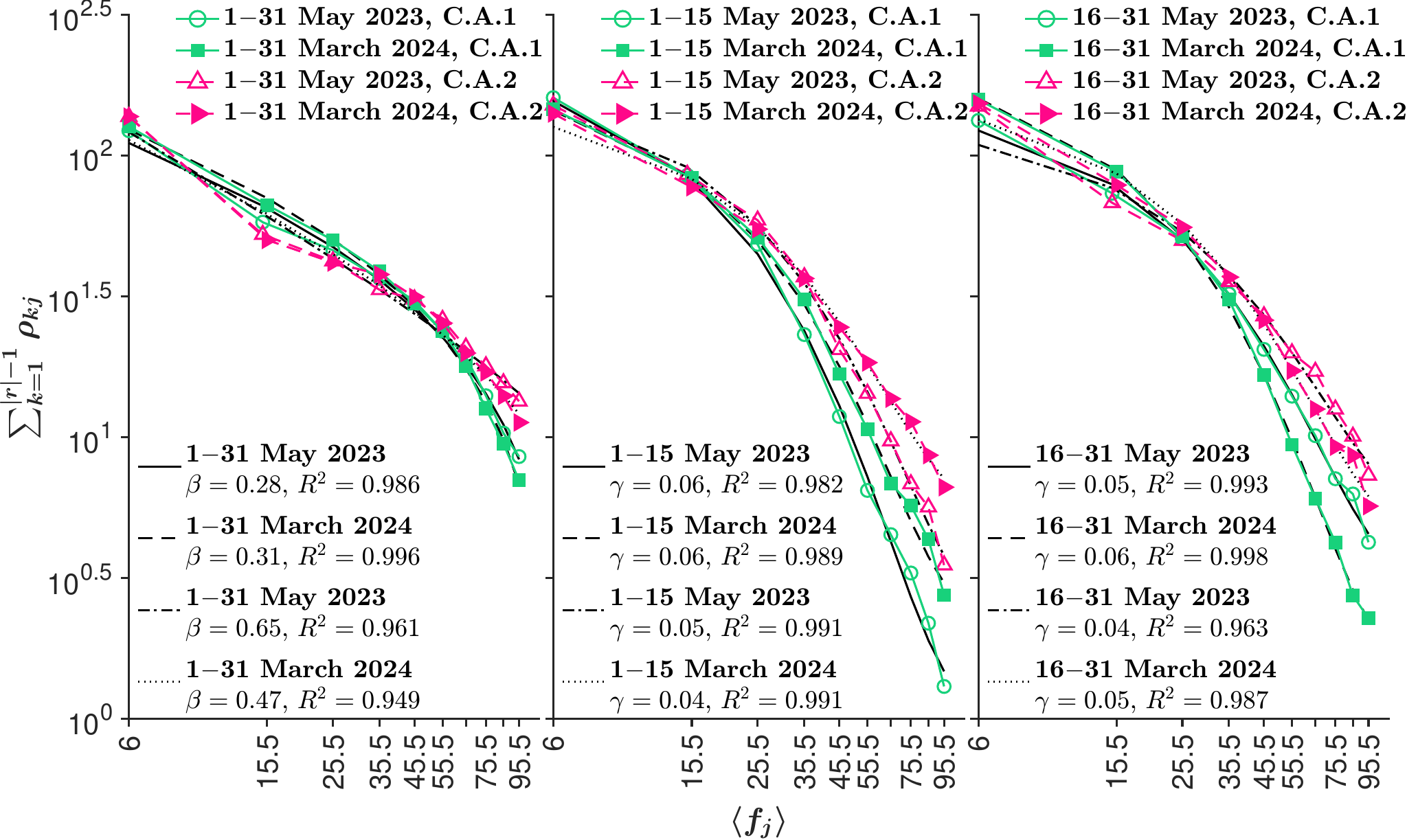}
    \caption{
    Aggregated visitor density (over distance) for frequency bin $\mathcal{F}_j$ as a function of mean visit frequency  $\langle f_j \rangle$.
    }
    \label{sum_rho_over_d_for_Fj}
\end{figure}

\begin{table}[H]
\centering
\caption{Model parameters for different time ranges and Census Areas (C.A.)}
\begin{tabularx}{\textwidth}{@{}l@{\hskip 1cm}l@{}}
\label{parameters_table_2}

\begin{tabular}{@{}lcccc@{}}
\toprule
\textbf{Dates} & \textbf{C.A.} & $\bm{a_2}$ & $\bm{\lambda}$ & \textbf{R\textsuperscript{2}} \\
\midrule
1—31 May 2023    & 1 & 11836.00 & $-$2.53 & 0.97 \\
1—15 May 2023    & 1 & 11969.00 & $-$2.71 & 0.98 \\
16—31 May 2023   & 1 & 19062.00 & $-$2.81 & 0.95 \\
1—31 March 2024  & 1 & 11415.00 & $-$2.52 & 0.97 \\
1—15 March 2024  & 1 & 15518.00 & $-$2.77 & 0.97 \\
16—31 March 2024 & 1 & 14370.00 & $-$2.71 & 0.97 \\
\bottomrule
\end{tabular}
&
\begin{tabular}{@{}lcccc@{}}
\toprule
\textbf{Dates} & \textbf{C.A.} & $\bm{a_2}$ & $\bm{\lambda}$ & \textbf{R\textsuperscript{2}} \\
\midrule
1—31 May 2023    & 2 & 7439.80 & $-$2.14 & 0.95 \\
1—15 May 2023    & 2 & 8282.00 & $-$2.29 & 0.96 \\
16—31 May 2023   & 2 & 7806.70 & $-$2.29 & 0.96 \\
1—31 March 2024  & 2 & 8091.20 & $-$2.12 & 0.95 \\
1—15 March 2024  & 2 & 9386.00 & $-$2.30 & 0.96 \\
16—31 March 2024 & 2 & 8677.40 & $-$2.24 & 0.95 \\
\bottomrule
\end{tabular}

\end{tabularx}
\end{table}
If we now look at the second marginalized quantity --- that is, the aggregated density (over distance) for frequency bin $\mathcal{F}_j$ --- we notice that it follows a mix of trends depending on the observation period, as illustrated in \cref{sum_rho_over_d_for_Fj}:

\begin{equation}
\sum_{k=1}^{|r|-1} \rho_{kj} = 
\begin{cases}
a_3 \cdot \langle f_j \rangle^{\mu} + a_4 & \text{for 1—31 May and March}, \\
a_3 \cdot \exp(\phi \cdot \langle f_j \rangle) + a_4 & \text{for 1—15 and 16—31 May and March},
\end{cases}
\end{equation}
When comparing the same observation periods across 2023 and 2024 within each Census Area (see \cref{parameters_table_3}), we observe that the exponential decay rates $\phi$ are approximately time- and space-invariant (within a tolerance of $\pm 0.02$). In contrast, the power-law exponents $\mu$ are time-invariant only in Census Area~1. Moreover, the values of $\mu$ are consistently steeper in Census Area~2 than in Census Area~1.

\begin{table}[H]
\centering
\caption{Model parameters for different time ranges and Census Areas (C.A.)}
\label{parameters_table_3}
\begin{subtable}[t]{\textwidth}
\centering
\begin{tabular}{lcccccc}
\toprule
\textbf{Dates} & \textbf{C.A.} & $\bm{a_3}$ & $\bm{\mu}$ & $\bm{\phi}$ & $\bm{a_4}$ & $\bm{R^2}$ \\
\midrule
1—31 May 2023    & 1 & 177.88 & $-$0.39 & —       & $-$25.38  & 0.98 \\
1—15 May 2023    & 1 & 121.30 & —       & $-$0.07 & 0.53  & 0.98 \\
16—31 May 2023   & 1 & 84.65  & —       & $-$0.05 & 1.52  & 0.99 \\
1—31 March 2024  & 1 & 206.09 & $-$0.41 & —       & $-$27.75  & 1.00 \\
1—15 March 2024  & 1 & 106.35 & —       & $-$0.06 & 1.08  & 0.99 \\
16—31 March 2024 & 1 & 120.57 & —       & $-$0.06 & 0.76  & 1.00 \\
\bottomrule
\end{tabular}
\end{subtable}

\vspace{1em}

\begin{subtable}[t]{\textwidth}
\centering
\begin{tabular}{lcccccc}
\toprule
\textbf{Dates} & \textbf{C.A.} & $\bm{a_3}$ & $\bm{\mu}$ & $\bm{\phi}$ & $\bm{a_4}$ & $\bm{R^2}$ \\
\midrule
1—31 May 2023    & 2 & 331.06 & $-$0.81 & —       & $-$0.78 & 0.96 \\
1—15 May 2023    & 2 & 106.91 & —       & $-$0.05 & 1.17  & 0.99 \\
16—31 May 2023   & 2 & 72.64  & —       & $-$0.04 & 2.70  & 0.95 \\
1—31 March 2024  & 2 & 261.15 & $-$0.67 & —       & $-$6.21 & 0.94 \\
1—15 March 2024  & 2 & 91.00  & —       & $-$0.05 & 2.70  & 0.98 \\
16—31 March 2024 & 2 & 97.45  & —       & $-$0.05 & 2.24  & 0.98 \\
\bottomrule
\end{tabular}
\end{subtable}
\end{table}
\noindent
The observed transition from an exponential decay in shorter temporal aggregated periods ($a_3 \cdot \exp(\phi \cdot \langle f_j \rangle) + a_4$) to a power-law decay in a longer aggregated period ($a_3 \cdot \langle f_j \rangle^{\mu} + a_4$) can be hypothesized as an emergent property stemming from the superposition of exponential distributions \cite{Priesemann2018,Mizzi2023}. To explore this hypothesis, we first define the decaying component of the aggregated density (over distance) for frequency bin $\mathcal{F}_j$, as $D(\langle f_j \rangle) \equiv \frac{1}{a_3} \left( \sum_{k=1}^{|r|-1} \rho_{kj} - a_4 \right)$, which is empirically equal to  $\langle f_j \rangle^{\mu}$ for the full-month period of observation. We then derive the power-law form from the superposition (or mixture) of exponential densities (see \cite{Feller1971}, page 438).
This derivation requires the assumption that the individual theoretical exponential decay rates, $\phi'$, which contribute to the aggregated longer-period function, follow a power-law distribution.
We can then write $P(\phi') = C \cdot (\phi')^{-\psi} \quad \text{for } \phi_{\min} \le \phi' \le \phi_{\max}$ as the probability density function of the individual exponential decay rates $\phi'$ that contribute to the longer-period aggregation, where $C$ is a normalization constant and $\psi$ is a positive exponent. The aggregated density $D(\langle f_j \rangle)$ is then given by the integral of these individual contributions weighted by their distribution $P(\phi')$, i.e. $D(\langle f_j \rangle) = \int_{\phi_{\min}}^{\phi_{\max}} (\phi' e^{-\phi' \langle f_j \rangle}) P(\phi') d\phi'$. The change of variable $u = \phi' \langle f_j \rangle$ (so that $d\phi' = \frac{du}{\langle f_j \rangle}$), leads us to 
$D(\langle f_j \rangle) = C (\langle f_j \rangle)^{-(2-\psi)} \int_{u_{\min}}^{u_{\max}} u^{(2-\psi)-1} e^{-u} du$.
In the approximation of $\phi'_{\min} \langle f_j \rangle \approx 0$ and $\phi'_{\max} \langle f_j \rangle \approx \infty$, the aggregated density $D(\langle f_j \rangle)$ takes on the form of a power law, $D(\langle f_j \rangle) \approx C \cdot \Gamma(2-\psi) \cdot (\langle f_j \rangle)^{-(2-\psi)}$, where the integral becomes the Gamma function. That approximation is valid under the condition that $2-\psi > 0$, or equivalently, $\psi < 2$.
Since the decaying component of the aggregated density, $D(\langle f_j \rangle)$, is empirically proportional to $\langle f_j \rangle^{\mu}$ over the full-month observation period, our derivation holds for $\mu = -(2-\psi)$. 
Analyzing the empirical $\mu$ values in \cref{parameters_table_3}, they yield to $\psi = \mu + 2$ values of $1.61$ (for 2023) and $1.59$ (for 2024) respectively, for Census Area 1, and $\psi$ values of $1.19$ (for 2023) and $1.33$ (for 2024) respectively, for Census Area 2. In all cases, the $\psi$ values are consistently less than $2$, thereby satisfying the condition required for the validity of the power-law approximation.

\section*{Discussion}
\label{Discussion}

With the intention of evaluating whether the general 
structure of 
\textcolor{black}{mobility flows}
persists under an extreme flooding event, we 
analyzed the corresponding visitor 
\textcolor{black}{densities}
using mobile phone data. 
These data were purchased from TIM, one of the largest Italian telecommunications providers,
\textcolor{black}{which held an overall market share of 27.5\% in Italy in March $2024$, including a 23.9\% share of the human SIM card market \cite{TIM_March2024}.}
Despite the fluctuations in the number of visitors across different census areas and observation periods, we found that the 
\textcolor{black}{distance-frequency distribution of visitors}
remained relatively stable. This suggests a robust scaling behavior that persists even during an extreme flood and closely resembles the structural form reported by Schläpfer et al. (2021), 
\textcolor{black}{despite differences in the underlying data aggregation.}
\textcolor{black}{In other words, this observed robustness indicates that, while local mobility patterns may be reorganized during a flood, the density of travelers continues to depend on distance and travel frequency, as it does in the absence of flooding. This insight should be incorporated into models of flood impacts as a fundamental assumption and boundary condition, and into the assessment of indirect economic damages, such as traffic disruptions.}
\\
\\
In addition to the visitor density, we derived two aggregated quantities as well, i.e. the  marginal density (over frequency) for distance bin $\mathcal{D}_k$ and the aggregated density (over distance) for frequency bin $\mathcal{F}_j$. These quantities allowed us to isolate and examine the individual effects of distance and visitation frequency on the 
\textcolor{black}{overall}
visitor density, which is inherently related to the product of distance and visit frequency. 
We found that the marginal density (over frequency) for distance bin $\mathcal{D}_k$ follows a power-law distribution, with an exponent that remains stable across both 2023 and 2024 when using monthly observation periods. Within each Census Area, this exponent shows only slight variations when shorter biweekly periods are considered.
Remarkably, although the exponent differs slightly between the two Census Areas, this variation cannot be attributed to the flooding event, as the exponent is time-invariant across the years --- that is, it remains constant both during the flooding period and nearly one year afterward.
In contrast, the aggregated density over distance, for frequency bin $\mathcal{F}_j$, exhibits a more complex behavior. 
Specifically, it follows an exponential decay during biweekly observation periods and a power-law decay when observed over a full-month period. We hypothesize that the power-law behavior observed in the full-month data emerges from the superposition of exponential distributions across the shorter timescales. Notably, the exponent of the exponential decay is both time- and space-invariant. Similarly, the power-law exponent appears to be time-invariant in Census Area 1 --- which was heavily affected by the flood --- but shows slight variations in Census Area 2. This suggests that the aggregated density over distance for frequency bin $\mathcal{F}_j$ is robust to the effects of the extreme flooding event. 
\\
\\
The persistence of scaling laws during the floods, despite infrastructure disruptions, potentially reflects enduring socio-economic needs such as access to essential services, work, and caregiving. Although our analysis does not directly examine individual motivations, the robustness of visitation patterns under extreme conditions is consistent with these factors.

\section*{Methods}
\label{Methods}

\subsection*{Basic definitions}
\label{Basic_definitions}

The number of 
\textcolor{black}{unique mobile users}
who travel to a census area of Faenza is stored in the matrix $V = (v_{ij})_{1 \leq i \leq m, 1 \leq j \leq n}$.
Each $i-$th row of $V$ corresponds to a distinct place of origin of the visitors, 
while each $j-$th column represents a specific frequency range of visits. Therefore, $v_{ij}$ denotes the number of 
\textcolor{black}{unique}
visitors 
\textcolor{black}{traveling}
from their $i-$th origin to a census area of Faenza, with a frequency of visit corresponding to the $j-$th frequency range. 
\textcolor{black}{The total number of unique users detected in a census area,
across all origins and frequency bins, is denoted by  $N$, and is computed as $N = \sum_{i}^{m}\sum_{j}^{n} v_{ij}$.}
The distance $d_i$ between each visitor's $i-$th place of origin and the census area is contained in the column vector $D = (d_i)$, where $i = 1, \ldots, m$ (the same index used to denote the $i-$th row of $V$). 
Regarding the frequencies, we do not have the exact visit frequency $\nu_j$ for each visitor, but rather frequency ranges $\mathcal{F}_j$, which are contained in the row vector $F = (\mathcal{F}_j)$, where $j = 1, \ldots, n$, corresponding to the $j-$th column of the matrix $V$. 
These frequency ranges are defined as disjoint intervals \( \mathcal{F}_j = \{ \nu_l \mid \nu_l \in [f_j, f_{j+1}) \} \), where \( j = 1, \dots, n \), and \( f = \{ 1, 11, 21, 31, 41, 51, 61, 71, 81, 91, 101 \} \) represents the edges of these intervals. The number of edges is $|f| = n + 1$, and the number of frequency bins is $|f| - 1$.
These frequency ranges, such as
$\mathcal{F}_1 = [1, 11), \mathcal{F}_2 = [11, 21), \dots, \mathcal{F}_{10} = [91, 101)$, are referred to as \quotes{frequency bins}.
To ensure consistency between the distance and frequency data, we sort the distances traveled by visitors into intervals, which we refer to as \quotes{distance bins}. These intervals are defined as $\mathcal{D}_k = \{ d_i \mid d_i \in [r_k, r_{k+1}) \}$, where $r = \{1, 11, 21, 31, 41, 51, 61, 71, 81, 91, 101\}$ are the edges of the bins, and $k = 1, \ldots, (|r|-1)$. The number of edges is $|r|$, and the number of distance bins is $|r| - 1$.
Since visitors can arrive at the census area from any direction, each edge $r_k$ of the distance bin can be interpreted as the radius of a circle centered at the census area's center. 
Consequently, the visitors coming from a specific distance bin are those whose origins lie within an annular region, where the inner radius is $r_k$ and the outer radius is $r_{k+1}$. 
Within this annular region, visitors are distributed in all directions around the census area.
\\
\\
We now introduce the density of visitors $\rho_{kj}$ \cite{Schläpfer2021} as the ratio of the number of visitors originating from the $k-$th distance bin $\mathcal{D}_k$, and whose visit frequency falls within the $j-$th frequency bin $\mathcal{F}_j$, to the area $A_k$ of the corresponding annulus:
\begin{equation}
    \rho_{kj} = \frac{\sum_{i \mid d_i \in [r_k, r_{k+1})} v_{ij}}{A_k}.
\end{equation}
Here, the summation is performed over all $i-$origins whose distance $d_i$ from the census area falls within the $k-$th distance bin $\mathcal{D}_k$.
The annulus area is calculated as $A_k = \pi r_{k+1}^2 - \pi r_k^2 = \pi (r_{k+1}^2 - r_k^2)$. 
By summing the density of visitors over all frequency bins, i.e. 
\begin{equation}
    \sum_{j=1}^n \rho_{kj} 
        = \sum_{j=1}^n \left( \frac{\sum_{i \mid d_i \in [r_k, r_{k+1})} v_{ij}}{A_k} \right)
        = \frac{\sum_{j=1}^n \sum_{i \mid d_i \in [r_k, r_{k+1})} v_{ij}}{A_k},
\end{equation}
we obtain the total density of visitors for a specific distance bin $\mathcal{D}_k$. This quantity, which we call the \quotes{marginal density (over frequency) for distance bin $\mathcal{D}_k$}, tells us how many visitors per unit area come from a particular annular region, regardless of whether they are frequent or infrequent visitors.
Conversely, we can also sum the visitor density over all distance bins, i.e. 
\begin{equation}
    \sum_{k=1}^{|r|-1} \rho_{kj}
        = \sum_{k=1}^{|r|-1} \left( \frac{\sum_{i \mid d_i \in [r_k, r_{k+1})} v_{ij}}{A_k} \right).
\end{equation}
We call this quantity the \quotes{aggregated density (over distance) for frequency bin $\mathcal{F}_j$}. This quantity provides a measure of the overall presence of visitors with a particular visit frequency, considering their origins across the entire set of distance bins. In other words, it highlights how prevalent a certain visit frequency is across all distance bins.

\subsection*{Adapting the distance–frequency scaling algorithm of Schläpfer et al. to the TIM dataset}
\label{Adaptation}

The Schläpfer et al. datasets \cite{Schläpfer2021} comprise a collection of $100 \times 100$ matrices, where each entry represents the number of visitors associated with a specific frequency and distance. The row index corresponds to the visit frequency, ranging from 1 to 100 visits per period, and the column index corresponds to the travel distance, ranging from 1 to 100 km. Leveraging this data structure, their algorithm \cite{Schläpfer2021} demonstrates the decay of the visitor density $\rho$ when the product among the annular radius, $r_q$, and the average frequency, $\langle f_p \rangle = \frac{f_p + f_{p+1}}{2}$, increases. 
To analyze this relationship, the algorithm first defines frequency bins. Within each bin, it iterates over the individual frequencies (i.e. over the rows of the matrix), extracting the corresponding arrays of visitor counts across all distances (i.e. across all the columns of the matrix). These arrays of visitor counts are then normalized by the area of the corresponding annular rings, $A_q = 2\pi r_q \cdot \delta r$, and subsequently averaged across all frequencies within the bin, 
obtaining a representative (average) visitor density, $\rho$, per unit frequency and per unit area, for each frequency bin, at each distance.
The algorithm adopts a step size of $\delta r = 1$ km, reflecting the unit increment in travel distances that follows the natural column-wise indexing of the matrix (i.e. $q = 1, 2, 3, \dots$). Finally, the code computes $\rho$ against $r_q \langle f_p \rangle$. 
\\
\\
The algorithm developed by Schläpfer et al. \cite{Schläpfer2021} is not directly applicable to the TIM dataset due to some differences. Specifically, TIM data does not include visitor counts for individual visit frequencies but rather provides aggregated counts within frequency ranges $\mathcal{F}_j$ (please note that the TIM matrices of visitor counts are transposed with respect to the  analogous matrices in Schläpfer et al. \cite{Schläpfer2021}, i.e. frequencies are associated with $p-$rows in the Schläpfer et al. matrices, and with $j-$columns in the TIM matrices).
Additionally, travel distances are provided as exact values, and not grouped with a step size $\delta r = 1$ km. 
It follows that these differences require an adaptation of the Schläpfer et al. algorithm \cite{Schläpfer2021}.
For each frequency bin (i.e. iterating through each $j-$column of the TIM matrices) we first extract the corresponding array of visitor counts 
and group them according to a step size of $\delta r = 1$ km. 
Next, we normalise the array of visitor counts by the area $A_k$ of the corresponding annular ring, yielding a density of visitors $\rho$ for that frequency bin, at each distance. 
Since the TIM visitor counts are aggregated over frequency ranges, rather than being available for individual frequencies, we further divide the resulting densities by the number of discrete frequencies in the frequency bin, given by $f_{j+1} - f_j$ (having unit-spaced integer frequencies). 
This step produces a representative visitor density $\rho_{kj}^{\mathrm{rep}}$ per unit frequency and per unit area, for each frequency bin at each distance, aligning our analysis with the approach used by Schläpfer et al. \cite{Schläpfer2021}.
Therefore, the density $\rho_{kj}^{\mathrm{rep}}$ is calculated as 
\begin{equation}
    \rho_{kj}^{\mathrm{rep}} = \frac{1}{f_{j+1} - f_j} \cdot \frac{\sum_{i \mid d_i \in [r_k, r_{k+1})} v_{ij}}{2\pi r_k \cdot \delta r}
\end{equation}
where $v_{ij}$ is the number of visitors from origin $i$ in frequency bin $\mathcal{F}_j$, $d_i$ is the distance from origin $i$, $[r_k, r_{k+1})$ is the $k$-th distance bin, $f_{j+1} - f_j$ is the width of the $j$-th frequency bin, and $\delta r = 1 \, \mathrm{km}$ is the annulus width.

\textcolor{black}{\subsection*{Minimum duration for detection of individuals}
\label{Minimum Stay Time}
The minimum stay time, denoted as $\tau$ in the Schläpfer et al. (2021) \cite{Schläpfer2021} study, refers to the minimum duration an individual must remain within a given spatial area for their presence to be counted, either as a visit or a transient passage.
The datasets provided by TIM, which pertain to the two Faenza census areas, did not include any minimum stay time filter, i.e., no time threshold was applied to define a visit.
As a result, all detections of individual presence — including both pass-throughs and extended visits — were included in the datasets regardless of their duration, starting from zero minutes, i.e. from $\tau = 0$. 
}

\section*{Acknowledgements}

S.L. expresses his gratitude to TIM for providing their data, 
and to Lei Dong for sharing a data sample from the Schläpfer et al. (2021) \cite{Schläpfer2021} study. 
\textcolor{black}{S.L. extends his appreciation to the Unione della Romagna Faentina and, in particular, to L. Angelini, A. Impellizzeri, and L. Marchetti (listed in alphabetical order) for helpful discussions and for providing the \quotes{Immediate Support Contribution} (CIS) data.}

\section*{Funding Declaration}
This work was partially self-funded by the first author, Simone Loreti.

\section*{Author contributions statement}
S.L. proposed and initiated the project, designed and performed the research, analyzed the data, formulated models and their underlying arguments, established the collaboration with TIM, and wrote the manuscript.
\textcolor{black}{M.S. discussed the results and revised the manuscript.}
\textcolor{black}{A.Z. performed geodata analysis, and reviewed the manuscript.}

\section*{Additional information}

\textbf{Competing interests:} The authors declare no competing interest.
\\
\\
\textbf{Costs for data:} Details regarding the costs of the TIM data are confidential and cannot be disclosed.
\\
\\
\textbf{Data and materials availability:} 
The anonymized TIM mobile phone datasets, together with supporting materials and documentation, will be made available on Zenodo, in compliance with TIM privacy and confidentiality policies. The CIS data, provided by the Unione della Romagna Faentina, cannot be publicly released due to privacy restrictions.

% ------ FIGURES -------------------------------------------------

\FloatBarrier

\bibliography{main}
\end{document}

\maketitle

%==================================================
\def\LT@output{%
  \ifnum\outputpenalty <-\@Mi
    \ifnum\outputpenalty > -\LT@end@pen
      \LT@err{floats and marginpars not allowed in a longtable}\@ehc
    \else
      \setbox\z@\vbox{\unvbox\@cclv}%
      \ifdim \ht\LT@lastfoot>\ht\LT@foot
        \dimen@\pagegoal
        \advance\dimen@-\ht\LT@lastfoot
        \ifdim\dimen@<\ht\z@
          \setbox\@cclv\vbox to \textheight{\unvbox\z@\copy\LT@foot}%
          \@makecol
          \@outputpage
          \setbox\z@\vbox{\box\LT@head}%
        \fi
      \fi
      \global\@colroom\@colht
      \global\vsize\@colht
      \vbox to \textheight
        {\unvbox\z@\box\ifvoid\LT@lastfoot\LT@foot\else\LT@lastfoot\fi}%
    \fi
  \else
    \setbox\@cclv\vbox to \textheight{\unvbox\@cclv\copy\LT@foot}%
    \@makecol
    \@outputpage
      \global\vsize\@colroom
    \copy\LT@head\nobreak
  \fi}
%==================================================

%==================================================
\NewDocumentCommand{\hoursbar}{ O{5cm} O{4pt} O{red} m m }{%
   \begin{tikzpicture}[baseline]
        \fill[#3] ({(#4/24)*#1},0) rectangle ({(#5/24)*#1},{#2});
        \draw (0,0) rectangle ({0.25*#1},{#2});
        \draw ({0.25*#1},0) rectangle ({0.5*#1},{#2});
        \draw ({0.5*#1},0) rectangle ({0.75*#1},{#2});
        \draw ({0.75*#1},0) rectangle ({#1},{#2});
   \end{tikzpicture}%
}
%==================================================

%==================================================
\def\pcbACSF#1{%% percent bar
   {\color[HTML]{000000}\rule{\fpeval{#1/\percentscale*\barwidth} cm}{\barheight}} #1
} %
\def\pcbBioM#1{%% percent bar
   {\color[HTML]{000000}\rule{\fpeval{#1/\percentscale*\barwidth} cm}{\barheight}} #1
} %
\def\pcbSeniorExcursion#1{%% percent bar
   {\color[HTML]{000000}\rule{\fpeval{#1/\percentscale*\barwidth} cm}{\barheight}} #1
} %
%==================================================

\newcommand{\barwidth}{5} %
\newcommand{\barheight}{4pt} %
\newcommand{\percentscale}{10000} %

\begin{figure*}
        \centering
        \caption{
        \textcolor{black}{Distributions of
        \textcolor{black}{visit}
        stay times, for two specific weeks in 2023:
        23$-$30 May 2023 (the last week of the flood) and 
        6$-$13 June (after the flood had receded), for both census areas of Faenza.
        The leftmost bars in all the distributions represent 
        \textcolor{black}{visits with stay times in the}
        0$-$15 minute
        \textcolor{black}{interval, denoted by $W_{1}$,}
        \textcolor{black}{corresponding to}
        individuals who passed through the census areas. 
        \textcolor{black}{The proportion of these short visits relative to the total number of visits across all stay-time intervals, $\sum_{s=1}^{S} W_s$ (where $s = 1, \ldots, S$), is approximately one-quarter of the total, specifically between 24\% and 28\%.
        Since a user can have multiple visits — likely with varying durations —, the total number of visits $\sum_{s=1}^{S} W_s$ is greater than or equal to the number of unique users $N$. Formally, $\sum_{s=1}^{S} W_s \geq N$.}
        }
        }
        \label{Supp_Fig_1}
        \includegraphics[width=\textwidth]{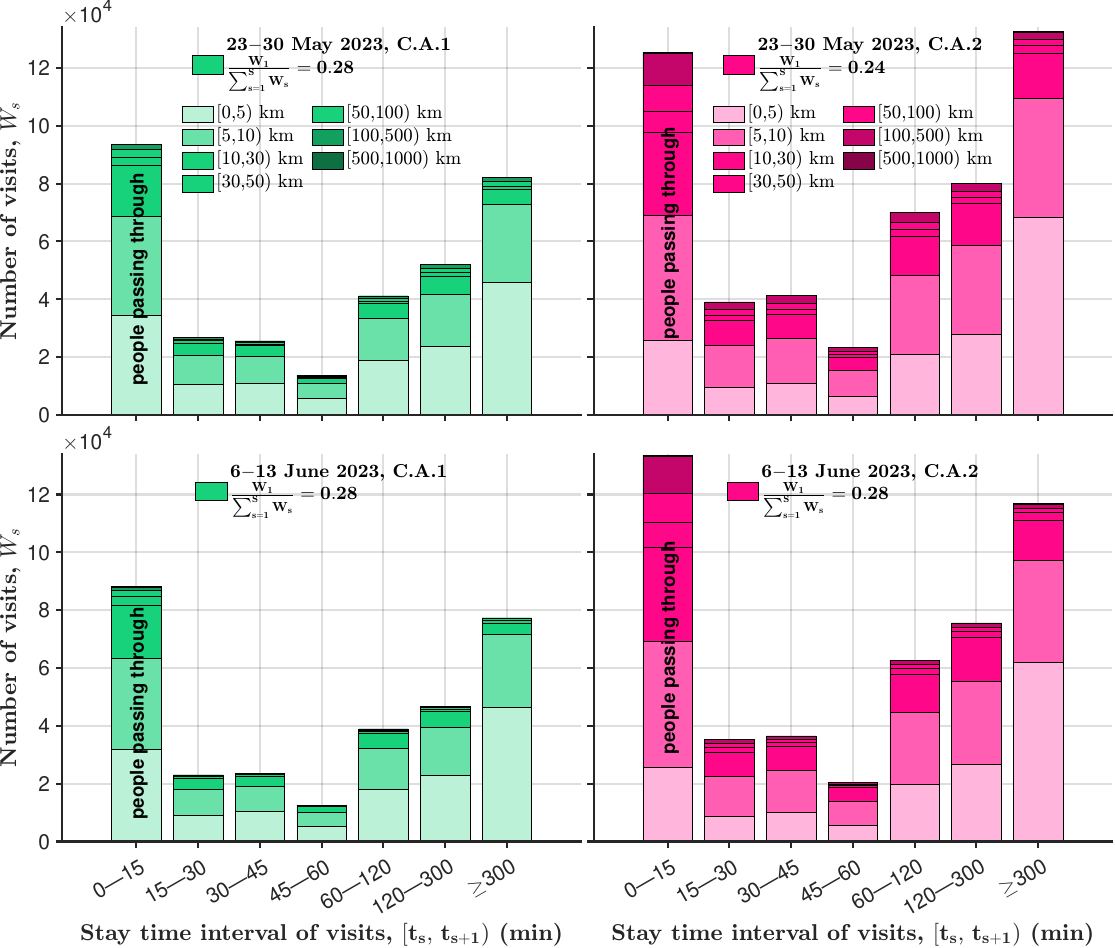}
\end{figure*}